\documentclass[12pt]{article}
\usepackage{graphicx}
\usepackage{amssymb}

\title{\bf Diffractive scattering on the deuteron}
\author{M.A.Braun\\
{\it S.Petersburg State University, Russia}}
\begin{document}

\maketitle
\input epsf

\def\beq{\begin{equation}}
\def\eeq{\end{equation}}
\def\ept{\epsilon_\perp}


\def\disc{{\rm Disc}}
\def\lra{\leftrightarrow}
\def\ep{\epsilon}
\def\eps{\varepsilon}
\def\aa{a^{(1)}}
\def\ab{a^{(2)}}
\def\qa{q_{1+}}
\def\qb{q_{2+}}
\def\ra{r_{1-}}
\def\rb{r_{2-}}
\def\bat{\bar{t}}
\def\ka{\kappa_+}
\def\bk{{\bf k}}
\def\bt{{\bf t}}
\def\bp{{\bf p}}
\def\bqa{{\bf q}_1}
\def\bqb{{\bf q}_2}
\def\bra{{\bf r}_1}
\def\brb{{\bf r}_2}
\def\tka{k_{1\perp}^2}
\def\tkb{k_{2\perp}^2}
\def\im{{\rm Im}\,}
\def\re{{\rm Re}\,}
\def\ci{{\rm ci}\,}
\def\tp{p_\perp^2}
\def\tpom{{\tilde P}}
\def\th{{\tilde h}}
\def\tv{{\tilde v}}
\def\ca{{\cal A}}

{\bf Abstract}

High-mass diffractive production of protons on the deuteron target is studied in the perturbative QCD in
the BFKL approach.
Leading order rearrangement contribution and the standard triple pomeron (the impulse approximation)
are studied. In the perturbative limit $\alpha_s\to 0$ the rearrangement contribution dominates.
Numerical estimates at realistic values of $\alpha_s$ and energies strongly depend on  assumptions
made about the behavior of the pomeron attached to the proton due to unitarization.
They indicate that irrespective of these assumptions in the realistic situation
the rearrangement and triple pomeron contributions
turn out to be of comparable magnitude  due to large dimensions of the deuteron.

\section{Introduction}
In the perturbative QCD collisions on heavy nuclear targets have long been the object
of extensive study. In the BFKL approach the structure function of DIS on a heavy nuclear target
is given by a sum of fan diagrams in which BFKL pomerons propagate and split by the
triple pomeron vertex ~\cite{bra1, blv1}. This sum satisfies the well-known Balitski-Kovchegov equation derived
earlier in different approaches ~\cite{bal,kov}. The corresponding inclusive cross-sections for gluon production
were derived in ~\cite{kov1,kov2}. Description of nucleus-nucleus collisions has met with less success.
For collision of two heavy nuclei in the framework of the Color Glass Condensate approach numerical
Monte Carlo methods were applied ~\cite{krasnitz,nara,lappi}. Analytical approaches
however have only given modest approximate results ~\cite{kovchegov, balitski,dusling}.
To understand the problem one of the authors (M.A.B)
turned to the simplest case of nucleus-nucleus interaction, namely the deuteron-deuteron collisions
~\cite{braun1,braun2}. It was found that in this case the diagrams which give the leading contribution are different from the
heavy nucleus case and include non-planar diagrams subdominant in $1/N_c$ where $N_c$ is the number of colors.

\begin{figure}
\begin{center}
\includegraphics[scale=0.6]{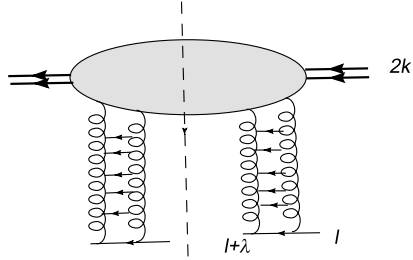}
\caption{Diffractive production by the deuteron}
\end{center}
\label{fig1}
\end{figure}

In this paper we continue our study of interactions with the deuteron target extending it to the high-mass
diffractive  production. Diffraction production of a heavy nucleus off the virtual photon  was studied
long ago ~\cite{kovlev} where the evolution equation was constructed for the cross-section integrated over all
variables of the produced nucleus. In our case we concentrate on the projectile rather than on the
diffractively produced object. We change the virtual photon to the deuteron and the heavy nucleus to the
proton with a given momentum.
The
diffractive production of protons by the deuteron
projectile with a large missing mass $M$ is illustrated in Fig.
\ref{fig1}. It is assumed that both $M$ and $s$ are large but
$M^2/s<<1$ so that the deuteron-pomeron amplitude can be given by the
pomeron exchanges.
In the BFKL, basically perturbative,  approach it is assumed that the  QCD coupling constant
$g$ is small but the overall rapidity $Y$ is large, so that the product $N_cg^2Y$ is of the order unity or larger.
In the BFKL approach one sums all powers of $N_cg^2Y$ considering $N_cg^2<<1$.
To classify contributions to the diffractive cross-section by their order of magnitude one has to decide whether
coupling of the BFKL pomeron to the proton carries a small $g^2$ or not. Modeling the proton by an "onium"
consisting of a quark-daiquiri pair at close distance between them (and so of large relative momentum) one may
think that the coupling is just $g^2$ and small. On the other hand the realistic proton does not contain large
relative momenta of its constituents on the average. Then one has no reason to ascribe any smallness to its
coupling to the pomeron.

Thus depending on whether we consider the protons on the average (case A) or their hard cores (case B) the order
of various contributions will be different.

\begin{figure}
\begin{center}
\includegraphics[scale=1.]{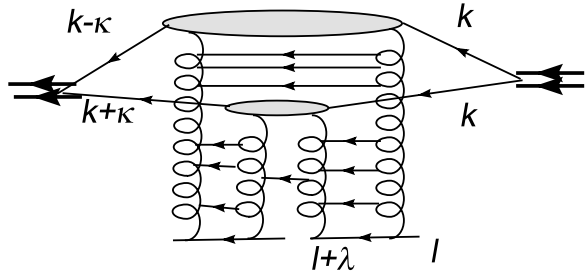}
\caption{Lowest order color rearrangement contribution}
\end{center}
\label{fig2}
\end{figure}

In case A one forgets about the couplings to the targets. Then
the leading contribution
 is  given by the color rearrangement diagram Fig. \ref{fig2}.

In the lowest order $N_c^2$ it does not involve any interactions of between the regions.
However this gives no contribution to the high-mass diffractive scattering. This contribution comes only in
the next order  $N_c^3g^2$:
Introducing new BFKL interaction between them will realize evolution in rapidity and provide additional
factors $(N_cg^2Y)^n$,
which, as mentioned, will not change the order of magnitude. Note that this contribution corresponds to double scattering
 and takes into account the deuteron structure

\begin{figure}
\begin{center}
\includegraphics[scale=1.0]{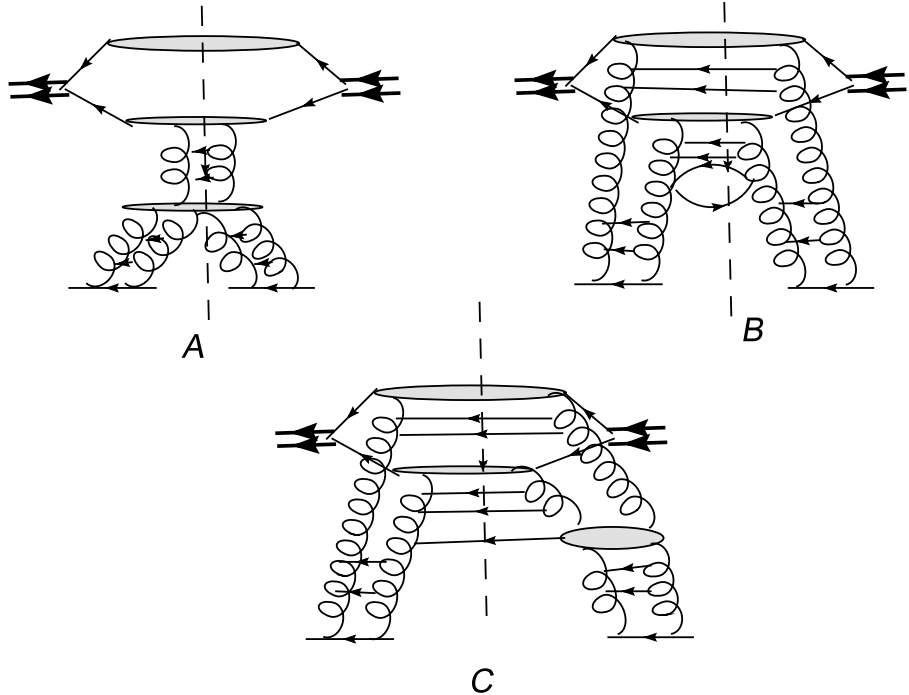}
\caption{Next-to-leading order contributions:
triple pomeron (A), corrections to the reggeon interaction (B),
RR$\to$RRP vertex (C).}
\end{center}
\label{fig3}
\end{figure}

Among the  subleading corrections we find, first of all, the expected
 diagram with the three-pomeron vertex \ref{fig3},A.
Its order is $N_c^4g^4$. So it is smaller that the rearrangement diagram in Fig. \ref{fig2}
by factor $N_cg^2$. However the same order of magnitude have the diagrams
with the first order correction to the rearrangement diagram \ref{fig3},B
and finally contribution from the RR$\to$RRP vertex \ref{fig3},C.
The first two corrections have a single scattering structure, whereas the last has a double scattering structure
as the leading rearrangement term
Fig. \ref{fig3},C.

These estimate are valid in case A when one forgets about the couplings to the proton.
In case B, when the proton is represented by its hard core, one has to take into account
couplings of the pomerons to the projectile. This gives  additional factors $g^2$ for single
scattering contributions,  Figs. \ref{fig3} A and B and factors $g^4$ for double scattering
contributions Figs. \ref{fig2} and \ref{fig3} C. As a result the tipple pomeron diagram and corrections to
the BFKL interaction become comparable to the rearrangement contribution.
The ratio of the formers to the latter is now $N_c^2g^2$, which may take any value depending on the relation
between $g$ and $1/N_c$. Still the contribution from Fig. \ref{fig3} C remains subdominant.

In this note we shall concentrate on the rearrangement term Fig. \ref{fig2}, which in any case gives a
substantial (leading in case A) contribution. The triple pomeron contribution is quite trivial and we
calculate it only to compare with the rearrangement term for realistic parameters and energies.
As to the rest of the subleading contribution we postpone their discussion for future
publications, since their calculation is far from straightforward and needs considerable efforts.

\begin{figure}
\begin{center}
\includegraphics[scale=0.6]{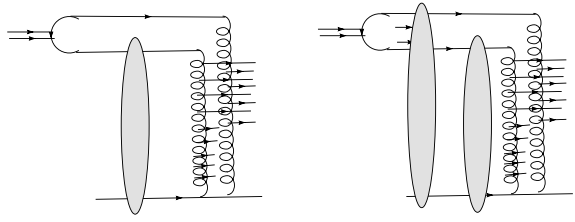}
\caption{Initial interactions of the participants}
\end{center}
\label{dif13}
\end{figure}

Note that, as is well known, the basic hard contributions we are going to discuss should be supplemented by
those coming from additional soft interactions
of the participants like shown in Fig. \ref{dif13} for production amplitudes. In the past they have been widely
discussed for
various diffraction processes. Their influence can be formulated by introduction of a certain  gap survival probability
factor $S^2$ which should multiply the hard contribution. This factor is obviously non-perturbative. For proton-proton
interactions this factor was calculated in ~\cite{khoze0,khoze1,khoze2,khoze4} in certain approximation schemes. It turned out to be small, of
order 0.1--0.2, and weakly falling with energy. Applied to our deutron case, in all probability, it  should be squared.
Then to pass to observables we can use the square of the gap survival probability factor $S^2$ from ~\cite{khoze2,khoze4}.

Generally the inclusive cross-section of the diffractive
proton production $d(2k)+p(l)\to p(l')+X$ is given by
\beq
I(l')\equiv\frac{(2\pi)^32l'_-d\sigma}{dl'_-d^2l'_\perp}=
\frac{1}{s}{\rm Im}{\cal A},
\label{eq1}
\eeq
where the forward amplitude ${\cal A}$ corresponds to Fig. \ref{fig1}.
Separating the deuteron lines we standardly find (see~\cite{bra1})
\beq
{\cal A}=\frac{1}{m}\int dz F(z)|\psi_d(r_\perp=0,z)|^2,
\label{eq2}
\eeq
where
\beq
F(z)=\frac{m}{k_+}\int\frac{d\kappa_+}{2\pi}H(\kappa_+)e^{izm\kappa_+/k_+}.
\label{ffun}
\eeq
Here $H$ is the high-energy part of ${\cal A}$,
$\kappa_+$ is the $+$-component of the transferred momentum $\kappa$
with all other components equal to zero.

For comparison, in the same process with a heavy nucleus projectile, the
contribution from the collision with two nucleons is given by (\ref{eq1}) with
\beq
{\cal A}=\frac{A(A-1)}{4m}\int d^2bdz_1dz_2 F(z_1-z_2)\rho({\bf b},z_1)\rho({\bf b},z_2),
\label{anuc}
\eeq
where $\rho({\bf b},z_1)$ is the nuclear density normalized   to unity.

The Glauber approximation corresponds to the contribution
which follows when $F(z)$ does not depend on $z$. Then the square of the deuteron
wave function converts into the average $<1/2\pi r^2>$ and in (\ref{anuc}) we find
integration over the impact parameter ${\bf b}$ of the square of the profile function
$T({\bf b})$. In standard cases the high-energy part contains $\delta(\kappa_+)$
\beq
H(\kappa_+)=2\pi \delta(\kappa_+)k_+D,\ \ {\rm so\ that}\ \
F=mD.
\label{defdd}
\eeq
Then for the deuteron
\beq
{\cal A}=D<1/2\pi r^2>_d
\label{ad}
\eeq
and for a large nucleus
\beq
{\cal A}=\frac{1}{4}A(A-1)D\int d^2bT^2(\bf b).
\label{aa}
\eeq

The final proton momentum is $l'=l+\lambda$. The missing mass is
$M^2=(2k-\lambda)^2=-4k_+\lambda_-$. So we find
\beq
\lambda_-=-\frac{M^2}{4k_+}=-\frac{M^2}{2s}k_+,\ \ \lambda_-<0.
\label{eq4}
\eeq
In the diffractive production, $M^2/s<<1$, so that
$|\lambda_-|<<l_-$ (in the c.m. system $k_+=l_-$).
The inclusive cross-section is then expressed via $M^2$ and $l'_\perp$
$l'_\perp$
\beq
I(M^2,l'_\perp)=\frac{(2\pi)^34sd\sigma}{dM^2d^2l'_\perp}.
\eeq
Passing to rapidity $y$ of the outgoing pomerons and $t=l_\perp^2$
we have
\beq
J(y,t)=\frac{d\sigma}{dy dt}=
\frac{M^2}{32\pi^2s^2}\im\ca,
\label{e5}
\eeq
where $M^2=M_0^2\exp(Y-y)$ and $M_0\sim 1$ GeV.

\section{The impulse approximation}
The impulse approximation for our process corresponds to Fig. \ref{fig3} A and is  the sum of
cross-sections off the proton and deuteron, each given by the triple pomeron contribution.
Although, as mentioned, for the deuteron projectile it may well be subleading, we present it here
because it is obviously expected from the start  and widely discussed. This cross-section is  a sum of contributions
from the proton and neutron components of the deuteron
\beq
J_{impulse}=J_p+J_n.
\label{impulse}
\eeq
Here for each contribution
\beq
J(y,t)=
\frac{N_c^4g^4}{4(2\pi)^7}\int \frac{d^2r_{12}d^2r_{23}}{r_{12}^2r_{23}^2r_{13}^2}
P_y(\lambda,r_{12})P_y(-\lambda,r_{23})e^{i\lambda r_{31}}
r_{13}^4\nabla_{13}^4P_{Y-y}(0,r_{13}).
\label{ip1}
\eeq
where $P_y(\lambda,r_{12})$ is the pomeron attached to the nucleon with the total transverse momentum $\lambda$ and
transverse distance between its reggeon components $r_{12}$.

For simplicity we concentrate on  proton emission in the forward direction, $\lambda_\perp=0$.
Then expression (\ref{ip1}) can be simplified by introducing
\[\delta^2(r_{12}+r_{23}+r_{31})=\frac{1}{(2\pi)^2}\int d^2qe^{i(r_{12}+r_{23}+r_{31})}.\]
Integrating over $r_{12}$, $r_{23}$ and $r_{31}$ we get
\beq
J_t(y,t=0)=\frac{N_c^4g^4}{2(2\pi)^7}\int d^2q\psi^2_y(q)\chi_{Y-y}(q),
\label{jtriple}
\eeq
where
\beq
\psi_y(q)=\int\frac{d^2r}{r^2} e^{iqr}P_y(r),\ \
\chi_{Y-y}(q)=\nabla_q^2 q^4 \nabla_q^2\psi_{Y-y}(q).
\label{psi}
\eeq
and all pomerons are taken in the forward direction.

\section{Leading order contribution}
\subsection{The rearrangement amplitude}
\begin{figure}
\begin{center}
\includegraphics[scale=1.0]{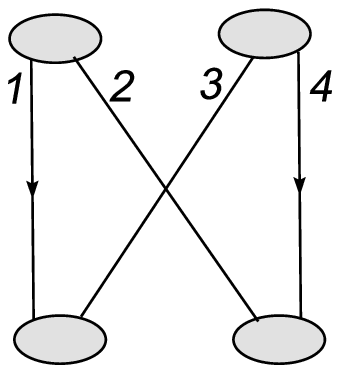}
\caption{Diagram with the redistribution of color and no interactions
 between
the pomerons of the projectile and target}
\label{fig60}
\end{center}
\end{figure}

\begin{figure}
\begin{center}
\includegraphics[scale=0.6]{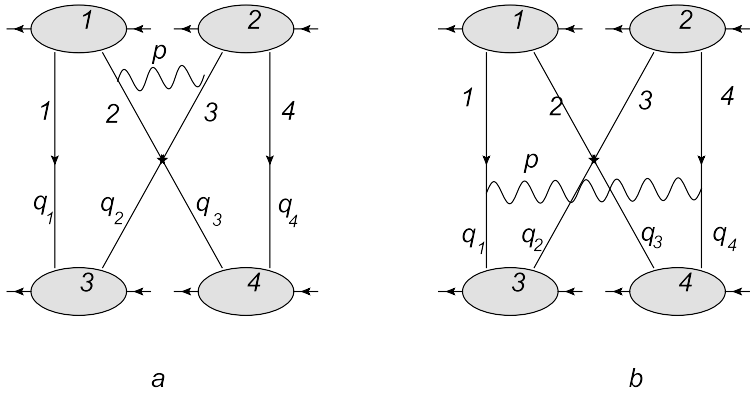}
\caption{Diagrams with the redistribution of color and one interaction between
the pomerons of the projectile and target}
\label{fig70}
\end{center}
\end{figure}

The leading order contribution corresponds to the diagram shown in Fig. \ref{fig2}.
It is given by a particular cut of the amplitude for the collision of the deuteron
with two targets, calculated in the forward direction in ~\cite{braun1}.
After cancelations of infrared divergent terms,  without energetic factors  and in the purely transverse form
the corresponding high-energy part is given by the sum of two terms
\beq
H_1=-i\frac{\partial}{\partial y}
\int_0^ydy'\int \frac{d^2q}{(2\pi)^2}\tpom^2_{y-y'}(q)\tpom^2_{y'}(q)
\label{fina1}
\eeq
and
\beq
H_2=-2i\int_o^ydy'\int\frac{d^2qd^2q'}{(2\pi)^4}\th(q,q'|q',q)
\tpom_{y-y'}(q)\tpom_{y-y'}(q')\tpom_{y'}(q)\tpom_{y'}(q'),
\label{fina2}
\eeq
which correspond to direct sewing of pomerons Fig. \ref{fig60} and one interaction between different pomerons,
Fig. \ref{fig70}. Here $\tpom_{y}(q)$ is the forward pomeron at rapidity $y$.

In  (\ref{fina1}) and (\ref{fina2}) both  the BFKL Hamiltonian $\th$ and pomerons $\tpom$ are taken in the
form symmetric respective to the initial and final states.
For the non-forward direction they are related to the standard Hamiltonian
$h$ and pomerons $P$ as
\beq
\tpom(q_1,q_2)=q_1q_2P(q_1,q_2),\ \
\th(q'_1,q'_2|q_1,q_2)=q'_1q'_2h(q'_1,q'_2|q_1,q_2)q_1^{-1}q_2^{-1}
\eeq
with
\beq
\th(q'_1,q'_2|q_1,q_2)=\tilde{v}(q'_1,q'_2|q_1,q_2)-(2\pi)^4\delta^2(q_1-q'_1)\delta^2(q_q-q_q')
\Big(\omega(q_1)+\omega(q_2)\Big),
\eeq
where $\omega(q)$ is the gluon Regge  trajectory and
the BFKL interaction is taken as
\beq
\tilde{v}(q'_1,q'_2|q_1,q_2)=\frac{g^2}{2\pi q_1q_2q'_1q'_2}
\Big(\frac{q_1^2{q'_2}^2+q_2^2{q'_1}^2}{(q_1-q'_1)^2}-(q_1+q_2)^2\Big).
\label{eq5}
\eeq
Here the momenta are transverse Euclidian , so that $q^2\equiv {\bf q}^2$.
As compared to ~\cite{braun1} we have added factor $-i$ corresponding
to transition from the $S$-matrix to the amplitude.

For our purpose we somewhat transform these expressions.
First we consider the corresponding non-forward expressions
providing each pomeron with its two momenta.
Next we perform the differentiation in (\ref{fina1}) to transform
\[
H_1=H_1^{(0)}+H_1^{(1)},\]
where
\beq
H_1^{(0)}=-i\int d\tau\tpom_{0}(4,1)\tpom_{0}(3,2)\tpom_{y}(4,3)\tpom_{y}(1,2)
\label{d10}
\eeq
and
\beq
 H_1^{(1)}=i\int_0^ydy'\int d\tau
\Big(\th_{41}+\th_{32}\Big)
\tpom_{y-y'}(4,1)\tpom_{y-y'}(3,2)\tpom_{y'}(4,3)\tpom_{y'}(1,2),
\label{d11}
\eeq
Here $\tau$ is the transverse phase volume (different in (\ref{d10}) and (\ref{d11})).
Notation $\tpom_{y}(1,2)$ means the pomeron at rapidity $y$ depending on two transverse
momenta of the reggeons $k_1$ and $k_2$.
In (\ref{d11}) it is understood that each Hamiltonian is to be applied to the
pomeron depending on the relevant momenta..

Taking part $H_2$. in the non-forward direction  we find the total
 transverse high energy part as
\[
 H^{tot}=H_1^{(0)}+H,
 \]
where
\[
H=-i\int_0^ydy'\int d\tau
\Big(\th_{13}+\th_{42}-\th_{41}-\th_{32}\Big)\]
\beq\times
\tpom_{y-y'}(4,1)\tpom_{y-y'}(3,2)\tpom_{y'}(4,3)\tpom_{y'}(1,2).
\label{dtot}
\eeq

\subsection{Leading order diffractive production}
\begin{figure}
\begin{center}
\includegraphics[scale=1.0]{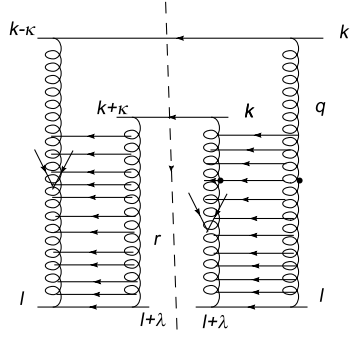}
\caption{Term $H_1^{(0)}$}
\label{fig51}
\end{center}
\end{figure}

We begin with term $H_1^{(0)}$, which graphically is illustrated in Fig. \ref{fig51}.
One observes that in the intermediate state we have only  contributions with small values of $M^2$
contained in $P_0(4,1)$ and $P_0(3,2)$. So this term does not give any contribution to diffractive production
at large $M^2$ and we are left with only the integral term in (\ref{dtot}).

\begin{figure}
\begin{center}
\includegraphics[scale=1.0]{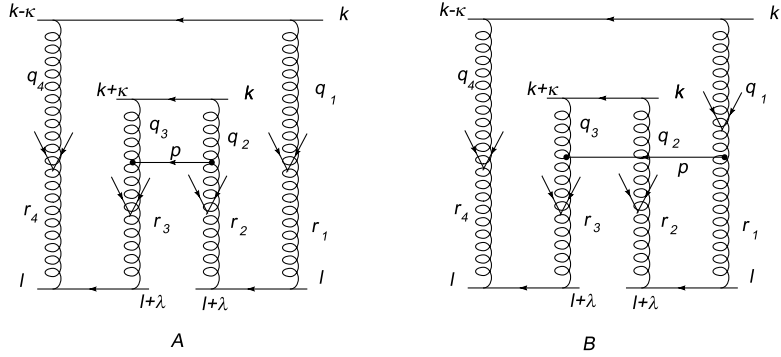}
\caption{LO contribution with terms $\tv_{23}$ (A) and  $\tv_{13}$ (B) in
(\ref{dtot1}) without evolution}
\label{fig50}
\end{center}
\end{figure}

To find the relevant energetic factors it will be necessary to restore the initial integrations
over the 4 momenta taking into account  the four impact factors of the pomerons in (\ref{dtot}).
For simplicity we choose   quarks for these impact factors,
remove evolution inside the pomerons. We also take into account both direct and crossed
contributions to the outgoing pomerons.
Then  we find for the transverse part (dropping the gluon trajectories in $\th$, canceled in the sum
of four $\th$)
\[
H_\perp=
-i\int_0^ydy'
\int d\tau
\Big(\tv_{13}+\tv_{24}-\tv_{14}-\tv_{23}\Big)\]\beq\times
\tpom_{y-y'}(4,1)\tpom_{y-y'}(3,2)\tpom_y'(4,3)\tpom_y'(1,2),
\label{dtot1}
\eeq
where we indicated by the  index $\perp$ that this is only the transverse part, which should be
multiplied by the appropriate energetic factor.
Terms with $\th_{23}$ and $\th_{13}$ are illustrated by diagrams A and B  in
Fig. \ref{fig50} respectively.

Consider the term with $\tv_{23}$ in  (\ref{dtot1}), shown in Fig\ref{fig50},A.
We have 6 transferred momenta $q_2$, $q_3$, $r_2$, $r_3$, $q_4=r_4$ and $q_1=r_1$
related by constraints
\beq
 \kappa=q_1+q_4=-q_3-q_2,\ \ \lambda=r_1+r_2=-r_3-r_4.
 \label{kinrel}
 \eeq
So we have two independent transferred momenta, for which we choose $q_2$ and $r_1$, with others
related to them as
$
q_3=-\kappa-q_2,\ \ r_2=\lambda-r_1,\ \ q_1=r_1,\ \ q_4=\kappa-r_1.
$

Let us study integrations over the 4 independent longitudinal momenta $q_{2\pm}$ and $r_{1\pm}$.
The 4 impact factors (with crossed and non-crossed reggeons) give
\[
256 k_+^4l_-^4(2\pi)^4\delta(2k_+q_{1-})\delta(2k_+q_2-)\delta(2l_-r_{1+})\delta(2l_-r_{4+})\]\beq
=(2\pi)^44s^2\delta(\kappa_+)\delta(r_{1-})\delta(r_{1+})\delta(q_{2-}).
\label{mshell}
\eeq

The four longitudinal integrations go over $q_{2\pm}$ and $r_{1\pm}$. Integration over
$q_+$ can be changed to that over $p_+$.
Integrations over $q_{2-}$ and $r_{1\pm}$ are lifted by the $\delta$ functions but the integration over
$p_+$ remains. In the diagram of Fig. \ref{fig50},A  its transversal part
$-iH_\perp^{(23)}$, which is just the term with $\th_{23}$ in (\ref{dtot1}, is multiplied by the propagator of the
intermediate gluon $-i/(p^2+i0)$. So the final longitudinal integration is
\[-i\int\frac{dp_+}{2\pi(2p_+p_-+p_\perp^2+i0)}=
-\frac{1}{4p_-}.\]
This brings us to the final energetic factor
\beq
-2\pi\delta(\kappa_+) 4s^2\frac{1}{4p_-}=2\pi\delta(\kappa_+) 4s^2\frac{1}{4\lambda_-}=
-2\pi\delta(\kappa_+) 4s^2 \frac{k_+}{M^2}
\label{enfac}
\eeq
and the  high-energy part corresponding to Fig. \ref{fig50},A will be
\beq
8\pi i\delta(\kappa_+) N_c^3g^2s^2 \frac{k_+}{M^2} H_\perp^{(23)}.
\label{h23}
\eeq
Now consider  the integration over $y$ in (\ref{dtot1}). Rapidity $y$ is expressed via the
missing mass $M^2$, which in turn is expressed via $l'_--l_=\lambda_-$:
\beq
y=\ln\frac{s}{M^2},\ \ M^2=-4k_+\lambda_-.
\label{rapidity}
\eeq
So we have
\[
\int dy=\int\frac{dM^2}{M^2}=\int \frac{dl'_-}{l'_-}\]
and we obtain (\ref{e5}) by removing integration over $y$ and fixing $y$
according to (\ref{rapidity}).

Next we study  the term with $\tv_{13}$ in (\ref{dtot1}), shown in Fig, \ref{fig50},B.
Here the 6 transferred momenta are $q_1$, $q_2$, $r_1$, $r_2$, $q_3=r_3$ and $q_4=r_4$,
constrained by conditions (\ref{kinrel}).
We take $q_1$ and $r_1$ as independent momenta. In terms of them
\beq
q_4=r_4=\kappa-q_1,\ \ q_2=r_2=\lambda-r_1,\ \ q_3=r_1-\lambda-\kappa,\ \
r_3=q_1-\lambda-\kappa.
\label{kinrel3}
\eeq
From impact factors(\ref{mshell}) together with (\ref{kinrel3}) we obtain a factor
\beq
(2\pi)^44s^2\delta(q_{1-})\delta(\lambda_--r_{1-})\delta(r_{1+})\delta(\kappa_+-q_{1+}).
\label{fac30}
 \eeq
Note that from (\ref{kinrel3}) it follows
\[ p_+=q_{1+},\ \ p_-=-r_{1-},\]
so that (\ref{fac30}) can be rewritten as
\beq
(2\pi)^44s^2\delta(q_{1-})\delta(\lambda_-+p_-)\delta(r_{1+})\delta(\kappa_+-p_+).
\label{fac31}
\eeq
After integration over $q_{1-}$, $r_{1+}$ and $p_\pm$ we find the transverse part $-iH_\perp^{(13)}$
multiplied by the propagator of the intermediate gluon $-i/(p^2+i0)$ in which the longitudinal
components of $p$ are fixed:
\[ p^2+i0=2p_+p_-+p_\perp^2+i0=-2\lambda_-\kappa_+ +p_\perp^2+i0.\]

Factor $-i/(p^2+i0)$ can be effectively transformed in a simpler expression if one takes into account that
it has to be eventually integrated over $\kappa_+$ with the weight $\exp (izm\kappa_+/k_+)$. At $k_+\to\infty$
we can neglect this weight to have the integral
\[
-i\int \frac{d\kappa+}{p^2+i0}=
\pi \frac{1}{2\lambda}\]
This is the same result that we would obtain if we substitute
\beq
-i\int\frac{dp_+dp_-}{p^2+i0}\delta(\kappa_+-p_+)\delta(\lambda_-+p_-)\to 2\pi\delta(\kappa_+)\frac{1}{4\lambda_-}
\eeq
As a result the corresponding energetic factor becomes identical to (\ref{enfac})
and the  high-energy part corresponding to Fig. \ref{fig50},C will be
\beq
8\pi i\delta(\kappa_+)N_c^3g^2 s^2 \frac{k_+}{M^2} H_\perp^{(13)}
\label{h12}
\eeq

The remaining interactions $\th_{14}$ and $\th_{24}$ in (\ref{dtot1})
can be studied in a similar manner. In fact the results can be achieved by the
interchange of reggeons 1234$\to$3412. So function $D$ in fact reduces to (\ref{dtot1}) with removed
integration over $y$.
Thus using the definition (\ref{defdd})
\[
D=4 i N_c^3g^2s^2 \frac{1}{M^2}
\int d\tau_\perp\Big(\tv_{13}+\tv_{24}-\tv_{23}-\tv_{14}\Big)\]\beq\times
\tpom(y-y',q_1,q_4)\tpom(y-y',q_2,q_3)\tpom(y',r_1,r_2)\tpom(y',r_3,r_4),
\label{dtotfin}
\eeq
where
\[y'=\ln\frac{s}{M^2},\ \ q_1+q_2=q_3+q_4=0,\ \ r_1+r_2=-r_3-r_4=\lambda\]
and all momenta are understood as purely transverse.
With the explicit expressions for $\tv_{ik}$ we get
\beq
D=8 i N_c^3g^2s^2\frac{1}{M^2}(T_A+T_B),
\eeq
where terms $T_A$ and $T_B$ correspond to Fig. \ref{fig50} A and B
\beq
T_A=\int\frac{d^2q_1d^2q_2}{(2\pi)^4}\,\frac{2q_1^6q_2^2}{(q_1+q_2)^2}P_{Y-y}(q_1)P_{Y-y}(q_2)
P^2_y(q_1)
\label{ta}
\eeq
and
\beq
T_B=-\int\frac{d^2q_1d^2q_2}{(2\pi)^4}q_1^2q_2^2\Big(\frac{(q_1^4+q_2^4)}{(q_1+q_2)^2}-
(q_1+q_2)^2\Big)P_{Y-y}(q_1)P_{Y-y}(q_2)
P_y(q_1)P_y(q_2).
\label{tb}
\eeq
Rewriting the two terms in $T_B$ as $T_B^{(1)}+T_B^{(2)}$, where
\beq
T_B^{(2)}=\int\frac{d^2q_1d^2q_2q_1^2q_2^2}{(2\pi)^4}
(q_1+q_2)^2
P_{Y-y}(q_1)P_{Y-y}(q_2)P_y(q_1)P_y(q_2)
\eeq
we get in the sum
\beq
T_A+T_B=
\int\frac{d^2q_1d^2q_2q_1^2q_2^2}{(2\pi)^4(q_1+q_2)^2}
P_{Y-y}(q_1)P_{Y-y}(q_2)\Big(P_y(q_1)-P_Y(q_2)\Big)\Big(q_1^4P_y(q_1)-q_2^4P_y(q_2)\Big)
+T_B^{(2)}.
\label{eqa54}
\eeq
As we observe the infrared singularity at $(q_1+q_2)^2=0$ is cancelled btween $T_A$ and $T_B$.
After angular integration we get the final cross-section in the forward direction
\[
J_r(y,t=0)=\frac{\alpha_sN^3}{4\pi^3}
\int_0^\infty dq_1dq_2q_1^3q_2^3P_{Y-y}(q_1)P_{Y-y}(q_2)\]\beq\times\Big[
\frac{1}{|q_1^2-q_2^2|}\Big(P_y(q_1)-P_Y(q_2)\Big)\Big(q_1^4P_y(q_1)-q_2^4P_y(q_2)\Big)
+(q_1^2+q_2^2)P_y(q_1)P_y(q_2)\Big].
\label{eqa55}
\eeq

\subsection{Evolution}
\begin{figure}
\begin{center}
\includegraphics[scale=0.6]{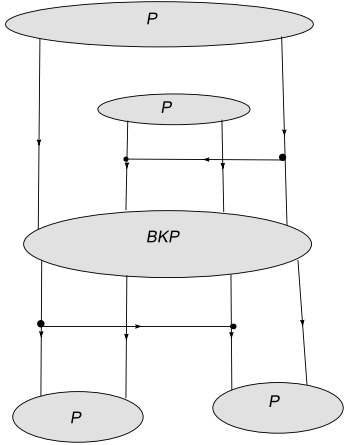}
\caption{Evolution with formation of the BKP state}
\label{fignew}
\end{center}
\end{figure}

Apart from the next-to-leading corrections to the found cross-section shown in Fig. \ref{fig3} new
contributions will be provided by evolution, that is by extra BFKL interactions among the reggeons.
Their immediate effect is to organize the fully-developed pomerons coupled to the projectiles and targets,
which actually has been already taken into account in our final formula (\ref{dtotfin}).
However evolution will also introduce additional contributions to the propagation of the four intermediate
reggeons between the projectiles and targets. In the high-color limit introduction of new BFKL interactions
between them will create the so-called BKP state, made of 4 reggeons, coupled to the projectiles and
targets by BFKL interactions necessary to transform their two-pomeron structure into an irreducible colorless
state, in which the reggeons are located on the cylinder surface. This contribution is schematically shown
in Fig. \ref{fignew}.
It is trivial to write the formal expression for it (see ~\cite{braun1}). However there
is not much use from it. On the one hand, the Green functions for the BKP states (except for the odderon)
are unknown and in all probability very complicated. On the other hand it is known that the BKP states grow much
slowlier with energy than the BFKL pomeron ~\cite{korch}. Therefore at high energy their  rapidity interval will be automatically squeezed
to finite rapidities, since the bulk of the contribution will come from the pomerons, which will occupy the whole
rapidity interval. Then one can hardly hope to have a small coupling constant inside the BKP state.
Within the BFKL approach with a fixed coupling constant adjusted to the overall rapidity interval this constant
will be small for the BKP state, so that one has to drop all extra interactions in it. This returns us to the
set of next-to-leading corrections in Fig. \ref{fig3}. Thus we do not see any necessity nor possibility to study
evolution between the projectiles and targets, at least until we know better the properties of the BKP state.

\section{Numerical estimates for the realistic situation}
The energy dependence of the cross-section is evidently determined by the behavior of
the pomerons attached to the participants. In the strict perturbative approach one takes them
to be the standard BFKL pomerons, which grow at large energies as $s^\Delta$ where
$\Delta=4(N_c\alpha_s/\pi)\ln 2$. Then at large $s$ the rearrangement contribution $J_r$ clearly dominates over
the triple pomeron one $J_t$ since
\beq
J_r\sim \alpha_s\frac{s^2}{M_0^4},\ \ J_t\sim\alpha_s^2\frac{s^2}{M^2M_0^2},\ \
\frac{J_t}{J_r}\sim\alpha_s\frac{M_0^2}{M^2},
\eeq
where one can take $M_0=1$ GeV and so $M_0^2<<M^2$. So not only the theoretical smallness of $\alpha_s$
but also the energy behavior make the triple pomeron contribution very small relatively.

Passing to concrete calculations we have first to couple the BFKL pomeron to the proton. To this aim we have to introduce
the proton dipole density in the momentum space $\rho(k)$ with the property $\rho(0)=0$.
We take
\beq
\rho(k)=\gamma k^2e^{-\beta k^2}.
\label {eqa42}
\eeq
The amputated pomeron $\phi_y(k)=k^2P_y(k)$ is then
\beq
\phi_y(k)=\int \frac{d^2k'}{(2\pi)^2}\rho(k')g_y(k',k)=
\frac{\gamma}{2\pi\beta k}\int d\nu  e^{y\omega(\nu)}k^{2i\nu}\beta^{i\nu}\Gamma(1-i\nu).
\label{eqa43}
\eeq
Here $g_y(k'k)$ is the BFKL Green function and $\omega(\nu)$ is the well-known BFKL eigenvalue. At small $\nu$
\beq
\omega(\nu)=\Delta-a\nu^2,\ \ \Delta=4\frac{N_c\alpha_s}{\pi}\ln 2,\ \ a=14\frac{N_c\alpha_s}{\pi}\zeta(3).
\eeq

To relate parameters $\gamma$ and $\beta$ to observables we calculate the proton-proton cross-section
\beq
\sigma(y)=
\frac{\gamma^2}{8\pi^2\beta^2}e^{y\Delta}\sqrt{\frac{\pi}{ay}}.
\label{eqa45}
\eeq
From this we can extract ratio $\gamma/\beta$ by comparison with the experimental data for $\sigma(y)$ at some
appropriate $y$. As to $\beta$ it is evidently related to the proton radius $R_p$, which we take to be 0.8 fm. We have
$\beta=R_p^2/4$. Both $\gamma$ and $\beta$ are dimensionful
\[{\rm dim}\,\alpha=-3,\ \ {\rm dim}\,\beta=-2.\]
In the asymptotic region at large $y$
\beq
P_y(k)=C_0 \frac{e^{y\Delta}}{\sqrt{y}}\frac{1}{k^3}\exp\Big(-\frac{\ln^2k^2}{4ya}\Big),
\ \ C_0=\frac{\gamma}{2\pi\beta}\sqrt{\frac{\pi}{a}}
\label{eqa46}
\eeq
and for pomerons $\psi$ and $\chi$ introduced by (\ref{psi}) we find in this limit (see Appendix 1.)
\[\psi_y(k)=\phi(k)=k^2P_y(k),\ \ \chi_y(k)=9\phi_y(k)\]
Note that the scale of $k$ is fixed  by the scale of $k'$ in
the integration with $\rho(k')$ and so with the scale of $\beta$. In the following we measure $\beta$ in mbn
and so $k^2$ in 1/mbn.

Using these asymptotic expressions we find the contribution from the triple pomeron
\beq
J_t^{(1)}(y,t0)=36\pi\alpha_s^2N_c^4C_0^3e^{(Y+y)\Delta}\frac{1}{y\sqrt{Y-y}}I_t^{(1)},
\label{eqa52}
\eeq
where
\beq
I_t^{(1)}=\int_0^\infty\frac{dq}{q^2}\exp\Big[-\frac{\ln^2q^2}{4a}\Big(\frac{2}{y}+\frac{1}{Y-y}\Big)\Big]
\label{eqa53}
\eeq
and from the rearrangement terms
\beq
J_r^{(1)}(y,t=0)=\frac{\alpha_sN_c^3}{4\pi^3}C_0^4<1/2\pi r^2>_de^{2Y\Delta}
\frac{1}{y(Y-y)}I_r^{(1)}
\label{eqa56}
\eeq
where
\[
I_r^{(1)}=\int_0^\infty\frac{dq_1dq_2}{q_1^3q_2^3}\exp\Big(-\frac{\ln^2q_1^2+\ln^2q_2^2}{4a(Y-y)}\Big)\]\[\times
\Big\{\frac{1}{|q_1^2-q_2^2|}\Big[q_2^3\exp\Big(-\frac{\ln^2q_1^2}{4ay)}\Big)-q_1^3\exp\Big(-\frac{\ln^2q_2^2}{4ay)}\Big)
\Big]\]\beq\times\Big[q_1\exp\Big(-\frac{\ln^2q_1^2}{4ay)}\Big)-q_2\exp\Big(-\frac{\ln^2q_2^2}{4ay)}\Big)\Big]
+(q_1^2+q_2^2)\exp\Big(-\frac{\ln^2q_1^2+\ln^2q_2^2}{4ay)}\Big)\Big\}.
\label{eqa57}
\eeq

Integrals $I_t^{(1)}$ and $I_r^{(1)}$ are convergent both in the ultraviolet and infrared. However in both
 $I_t^{(1)}$ and
especially $I_r^{(1)}$ the bulk of the contribution comes from extremely low values of $q$, where convergence is achieved due
to the damping exponentials $\exp(-c\ln^2q^2)$. As a result the cross-sections turn out to be absurdly large, of order
10$^{10}$ bn/Gev$^2$.
The BFKL approach is certainly not valid in this region. So to be closer to reality we cut the integrations
at values $q< \Lambda_{QCD}\sim 0.3$ GeV. We also somewhat diminish the BFKL intercept $\Delta$ to make it more compatible
with the data. We choose $\Delta=0.12$ in the hope that unitarity corrections will reduce it to this admissible value.
For hard interactions we take $\alpha_s=0.2$ and naturally $N_c=3$. For the deuteron, using the Hulthen wave function, we find
\beq
<1/2\pi r^2>_d=0.0764\ \ 1/{\rm fm}^2
\eeq

\begin{figure}
\begin{center}
\includegraphics[scale=0.6]{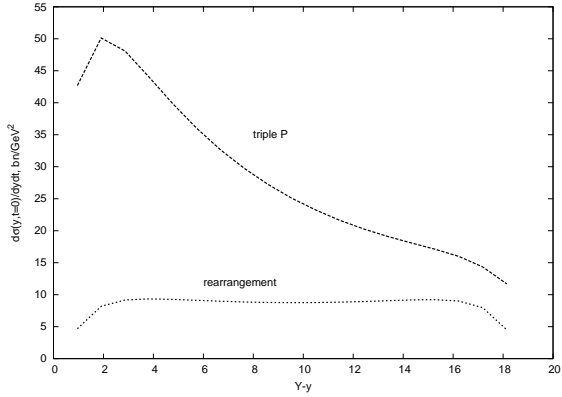}
\caption{Diffractive cross-section in the perturbative approach
in bn/GeV$^2$. The overall rapidity is $Y=19.1$}
\label{pcrsec}
\end{center}
\end{figure}

The calculated in this manner cross-sections at $Y=19.1$ corresponding to energy 14 TeV are
illustrated in Fig. \ref{pcrsec} as a function of $Y-y$. We recall that
the missing mass squared $M^2=\exp(Y-y)$ Gev$^2$. As we see the rearrangement cross-section is somewhat smaller than
the triple pomeron contribution due to very low value of $<1/2\pi r^2>_d$. But then the relation between them is very
sensitive to the infrared cut: the rearrangement part grows much faster with its lowering.

\begin{figure}
\begin{center}
\includegraphics[scale=0.6]{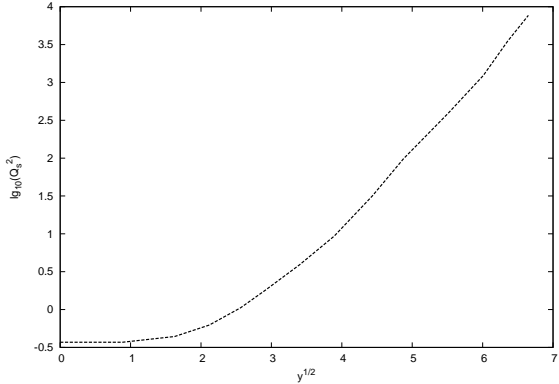}
\caption{$Q_s^2(y)$ for the proton}
\label{qsat}
\end{center}
\end{figure}

Still the  behavior of the pomerons with all unitarity corrections included should be
seriously different from the pure BFKL pomeron, both in the region of high energies and especially of
low momenta, where we expect the phenomenon of gluon saturation to take place.
So, as an alternative,
we shall use
expressions for the pomerons based on the latter phenomenon. Prompted by the approximate
form for the developed unintegrated gluon densities resulting from the Balitski-Kovchegov evolution equation we
take in the coordinate space for the pomeron attached to the proton
\beq
P_y(r)=\frac{2\pi}{g^2}S_\perp\Big(1-e^{-Q_s^2(y)}\Big).
\label{pom}
\eeq
Here $Q_s(y)$ is the proton saturation  momentum. Its $y$-dependence was presented in ~\cite{dusling}
and is shown in Fig. \ref{qsat}. Factor $S_\perp$ is the transverse area of the proton. It appears because the
standard unintegrated gluon density is calculated per unit of the transverse area of the target.
Factor $2\pi/g^2$ is due to different normalization of  the unintegrated gluon density and the BFKL
pomeron ~\cite{BLV}

With the form (\ref{pom}) both $\psi$ and $\chi$ can be found analytically. If
we define
\beq
x=\frac{q}{2Q_s(Y-y)},\ \ \rho=\frac{Q_s^2(Y-y)}{Q_s^2(y)}
\label{defxr}
\eeq
then we find (see Appendix 2. for details)
\beq
\psi(q)=-\frac{2\pi^2}{g^2}S_\perp{\rm Ei}(-\rho x^2),\ \ \chi(q)=64x^2(x^2-1)e^{-x^2}.
\eeq

As a result we get the cross-section from the triple pomeron
\beq
J^{(2)}_{t}(y,t=0)=\frac{N_c^4}{16\pi\alpha_s}S_\perp^3Q_s^2(Y-y)I_t^{(2)}(\rho),
\label{j3p}
\eeq
where
\beq
I_t^{(2)}(\rho)=64\int_0^\infty dzz(1-z)e^{-z}{\rm Ei}^2(-\rho z).
\label{jt2}
\eeq

Note that due to operator $\nabla_q^2 q^4 \nabla_q^2$ function $\chi(Y-y)$ is not positive
for all values of $0<y<Y$ but rather only at certain distance of its ends. Closer to $0$ or $Y$
it becomes negative and pathological (either close to zero or to $-\infty$). This property is apparently
the consequence of our choice for the pomeron wave function, which is not conformal invariant, unlike
the perturbative BFKL pomeron, for which the above operator is harmless. In the following
we exclude from consideration the intervals in $Y-y$ for which $\chi$ is negative.

To calculate the rearrangement contribution
(\ref{dtotfin})   we use according to (\ref{pom})
\beq
P_y(q)=\frac{2\pi}{g^2}S_\perp\Big((2\pi)^2\delta(q)-\frac{\pi}{Q_s^2(y)}e^{-q^2/4Q^2_s(y)}\Big).
\label{pomm}
\eeq
Due to factors $q_1^2$ and $q_2^2$ in (\ref{eqa55}) the $\delta$-terms in (\ref{pomm}) give no contribution.
So one obtains
\beq
D=\frac{s^2}{M^2}\frac{16(2\pi)^3N_c^3}{\alpha_s^3}S_\perp^4Q_S^2(Y-y)I_r^{(2)}(\rho),
\label{dr2}
\eeq
where
\[
I_r^{(2)}(\rho)=\rho^2\int_0^\infty x_1dx_1e^{-(1+\rho)x_1}\int_0^{x_1} x_2dx_2e^{-(1+\rho)x_2}\]\beq\times
\Big\{\frac{2x_2^2}{x_1-x_2}\cosh\Big(\rho(x_1-x_2)-1\Big)
+(x_1+x_2)e^{-\rho(x_1-x_2)}\Big\}.
\label{i2}
\eeq
The cross-section is then
\beq
J_{r}^{(2)}(y,t=0)=\frac{4\pi N_c^3}
{\alpha_s^3}S_\perp^4Q_s^2(Y-y)I_r^{(2)}(\rho)<1/2\pi r^2>_d.
\label{jr}
\eeq

Before any calculations one has the ratio
\beq
\frac{J_r}{J_t}=\frac{9N_c\alpha_s^2}{64\pi^2 S_\perp<1/2\pi r^2>_d}\,\frac{I_1(\rho)}{I_2(\rho)}.
\label{ratio}
\eeq
One observes that for very small $\alpha_s$ the rearrangement contribution clearly dominates.
However with realistic values of $\alpha_s$ and $N_c$ the situation changes.
Due to the large deuteron dimension, on the one hand, and the relation $I_1>>I_2$ for realistic rapidities,
on the other,
the ratio becomes around 10\%.

\begin{figure}
\begin{center}
\includegraphics[scale=0.6]{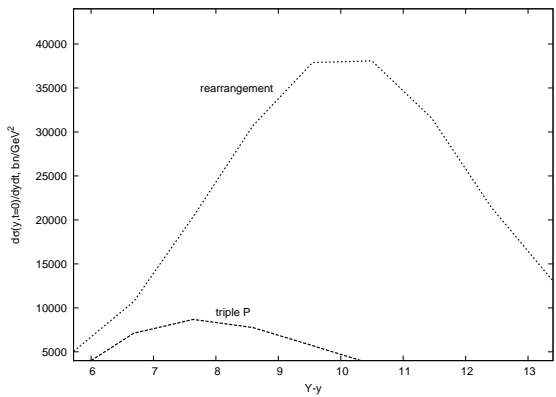}
\caption{
Diffractive cross sections with gluon saturation in
bn/GeV$^2$. The overall rapidity is $Y=19.1$}
\label{crsec}
\end{center}
\end{figure}

The   cross-sections from the triple pomeron and rearrangement calculated in this approach  are shown in
 Fig. \ref{crsec} for different values of $Y-y$ in bn/GeV$^2$.

\section{Discussion}
We have studied the high-mass diffractive proton production off the deuteron.
Our attention has been concentrated on the contribution from the color rearrangement diagram,
which should dominate the cross-section in the strict perturbative approach. We have derived the corresponding
cross-section and demonstarted its infrared finiteness. To compare we also
have included the obvious impulse approximation contribution, that is the sum of cross-sections
off the proton and neutron with the triple pomeron interaction.

As expected the results crucially depend on the unknown properties of the pomeron coupled to the proton,
modified by all sorts of unitarity corrections. With minimal modifications including lowering of the
intercept $\Delta$ and cutting in the infrared at momenta of the order $\Lambda_{QCD}$ the results are
presented in Fig. \ref{pcrsec}. More drastic modifications taking into account gluon saturation at low momenta
give cross-sections shown in Fig. \ref{crsec}. The results from these two choices are very different in their
magnitude, $M^2$ -dependence and the relation between the triple pomeron and rearrangement contributions.
One hopes that experimental studies may decide for the better choice and thus tell us something
on the behavior of the pomeron coupled to the proton. We recall that observable cross-section are to be obtained
from ours after multiplication by the squre of survival gap probability factor $S^2$ borrowed from
~\cite{khoze2,khoze4}. This will diminish our cross-sections by two orders of magnitude.
The main message we can extract from our calculation is that  in fact both triple pomeron and rearrangement
term give comparable contribution at the LHC energies with a realistic value of the coupling constant for
hard processes.

Next step is to take into account, first, evolution between pomerons attached to projectiles and targets
(Fig. \ref{fignew}) and, second, higher order corrections indicated in Fig. \ref{fig3} B and C.
Again in the purely perturbative approach they should be small. But for realistic parameters and
energies this may be not so. However calculation of  these corrections is apparently a highly
complicated task and so will be postponed for future investigation.

\section{Acknowledgements}
This work has been supported by the S.Pertersburg University grant 11.38.223.2015 and RFBR grant 15-02-02097

\section{Appendix 1. BFKL pomerons}
Elementary eigenfunctions of the BFKL Hamiltonian in the forward direction are the semi-amputated
pomerons
\beq
\phi_\nu(k)=\sqrt{2}k^{-1+2i\nu}
\eeq
normalized according to
\beq
\int\frac{d^2k}{(2\pi)^2}\phi_{\nu'}^*(k)\phi_\nu(k)==\delta(\nu-\nu')
\eeq
So the Green function is
\beq
g_y(k',k)=\frac{2}{kk'}\int d\nu e^{y\omega(\nu)}\Big(\frac{k}{k'}\Big)^{2i\nu}
\label{eqa41}
\eeq
where at small $\nu$
\beq
\omega(\nu)=\Delta-a\nu^2,\ \ \Delta=4\frac{N_c\alpha_s}{\pi}\ln 2,\ \ a=14\frac{N_c\alpha_s}{\pi}\zeta(3)
\eeq

For the triple pomeron contribution we have to know two other pomerons determined via the pomeron $P(r)$ in
the coordinate space. First
\beq
\psi_y(q)=\int \frac{d^2r}{r^2}P_y(r)e^{1kr}
\label{eqa47}
\eeq
To find it we note that
\beq
k^2\nabla_k\psi_y(k)=k^2P_y(k)=\phi_y(k)
\label{eqa48}
\eeq
In the $\nu$ representation
the $k$ dependence of $\psi_y(k)$ is the same as of $\phi_y(k)$. So we seek
\beq
\psi_\nu(k)=d_\nu k^{-1+2i\nu},\ \ k^2\nabla_k^2\psi_\nu(k)=(1-2i\nu)^2d_\nu k^{-1+2i\nu}
\eeq
From (\ref{eqa48}) then
\[
\psi_\nu(k)=\frac{\phi_\nu(k)}{(1-2i\nu)^2}\]
so that
\beq
\psi_y(k)
=
\frac{\alpha}{2\pi\beta k}\int d\nu  e^{y\omega(\nu)}k^{2i\nu}\frac{\beta^{i\nu}\Gamma(1-i\nu)}{(1-2i\nu)^2}
\label{eqa49}
\eeq

Finally we need
\[\chi_y(r)=r^4\nabla^4r^{-2}P_y(r)=r^4\nabla^4\int \frac{d^k}{(2\pi)^2}e^{ikr}\psi(k)=
\nabla_k^4k^4\psi_y(k)\]
In the $\nu$ representation
\[\nabla^4k^4k^{-1+2i\nu}=(3+2i\nu)^2(1+2i\nu)^2k^{-1+2i\nu}\]
so that
\beq
\chi_y(k)
=
\frac{\alpha}{2\pi\beta k}\int d\nu  e^{y\omega(\nu)}k^{2i\nu}(3+2i\nu)^2(1+2i\nu)^2
\frac{\beta^{i\nu}\Gamma(1-i\nu)}{(1-2i\nu)^2}
\label{eqa50}
\eeq

In the asymptotic region at large $y$ contributions come from $\nu<<1$ so that we can neglect $\nu$
in the additional factors in (\ref{eqa49}) and (\ref{eqa50}). Then we get a simple result
\beq
\psi_y(k)=\phi_y(k),\ \ \chi_y(k)=9\phi_y(k)
\label{eqa51}
\eeq

\section{Appendix 2. Functions $\psi(q)$ and $\chi(q)$ with the pomeron (\ref{pom}) }

With the expression (\ref{pom}) for the pomeron in the configuration space the semi-amputated
momentum space pomeron $\psi$ can easily be found analytically. We have
\beq
\psi(q)=\frac{2\pi}{g^2}S_\perp\int \frac{d^2r}{r^2}e^{iqr}\Big(1-e^{-Q^2r^2}\Big)=
\frac{(2\pi)^2}{g^2}S_\perp\int_0^\infty \frac{dr}{r}{\rm J}_0(qr)\Big(1-e^{-Q^2r^2}\Big).
\label{phi1}
\eeq
To avoid dealing with infrared divergent expressions we consider the integral as a
limit
\beq
\lim_{p\to 0}
\int_0^\infty dr r^{p-1}{\rm J}_p(qr)\Big(1-e^{-Q^2r^2}\Big)=
\lim_{p\to 0}\Big(I_1-I_2\Big).
\label{phi20}
\eeq
At finite positive $p$ both $I_1$ and $I_2$ are known ~\cite{ryg,prud}.
\beq
I_1=\int_0^\infty r^{p-1}J_p(qr)\Big(1-e^{-Q^2r^2}\Big)=\frac{2^{p-1}\Gamma(p)}{q^p},
\label{i1a}
\eeq
\beq
I_2=\frac{2^{p-1}}{q^p p}\Big(\frac{q^2}{4Q^2}\Big)^p\, _1{\rm F}_1\Big(p,p+1.-\frac{q^2}{4Q^2}\Big).
\label{i2a}
\eeq
The divergent terms at $p\to 0$ cancel in the difference $I_1-I_2$. So we need to know terms
linear in $p$ the expression
\[\Gamma(p+1)-\Big(\frac{q^2}{4Q^2}\Big)^p\, _1{\rm F}_1\Big(p,p+1.-\frac{q^2}{4Q^2}\Big).\]
One has
\[\Gamma(p+1)=1-pC_E,\]
where $C_E$ is the Eiler constant. Then
\[\Big(\frac{q^2}{4Q^2}\Big)^p=1+p\ln\frac{q^2}{4Q^2}\]
and
\[_1{\rm F}_1\Big(p,p+1.-\frac{q^2}{4Q^2}\Big)=1+p\Big[{\rm Ei}\Big(-\frac{q^2}{4Q^2}\Big)-C_E
-\ln\frac{q^2}{4Q^2}\Big].\]
Collecting all terms we find
\[\lim_{p\to 0}\Big(I_1-I_2\Big)=-\frac{1}{2}{\rm Ei}\Big(-\frac{q^2}{4Q^2}\Big),\]
so that finally
\beq
\psi(q)=-\frac{2\pi^2}{g^2}S_\perp{\rm Ei}\Big(-\frac{q^2}{4Q^2}\Big).
\label{phi2}
\eeq

To find $\chi(q)$ we have to know operator
\[\hat{Z}=\nabla_q^2 q^4\nabla_q^2,\ \
\nabla_q^2=\frac{\partial^2}{\partial q^2}+\frac{1}{q}\,\frac{\partial}{\partial q}.\]
Trivial calculations give
\beq
\hat{Z}=x^4\Big(\frac{\partial}{\partial x}\Big)^4+10x^3\Big(\frac{\partial}{\partial x}\Big)^3
+23x^2\Big(\frac{\partial}{\partial x}\Big)^2+9x\frac{\partial}{\partial x},\ \
x=\frac{q}{2Q(Y-y)}.
\label{z}
\eeq
Action of this operator on $\phi(q)$ can be found by the following relations.
If $z\equiv -x^2$ then
\[ \frac{\partial}{\partial x}{\rm Ei}(-x^2)=\frac{2}{x}e^z,\ \
\Big(\frac{\partial}{\partial x}\Big)^2{\rm Ei}(-x^2)=-2e^z\Big(\frac{1}{x^2}+2\Big),\]
\[ \Big(\frac{\partial}{\partial x}\Big)^3{\rm Ei}(-x^2)=
4e^z\Big(\frac{1}{x^3}+\frac{1}{x}+2x\Big),\ \
 \Big(\frac{\partial}{\partial x}\Big)^4{\rm Ei}(-x^2)=
-4e^z\Big(\frac{3}{x^4}+\frac{3}{x^2}+4x^2\Big).\]
As a result
\beq
\hat{Z}{\rm Ei}(-x^2)=64x^2(x^2-1)e^{-x^2}.
\label{hz}
\eeq

\end{document}